\documentclass[10pt, draft]{article}
\usepackage{amsfonts}
\usepackage{amssymb}
\usepackage{amsthm}
\usepackage{amsmath} 
\usepackage{graphicx}
\usepackage{color}
\date{}
\newcommand{\sss}{\setcounter{equation}{0}}
\newtheorem{theorem}{THEOREM}[section]

\newtheorem{remark}[theorem]{REMARK}

\newcommand\beq{\begin{equation}}
\newcommand\ene{\end{equation}}

\begin{document}
\baselineskip=20 pt
\parskip 6 pt

\title{The number of eigenvalues of  the  matrix Schr\"odinger operator on the half line with general boundary conditions
\thanks{ PACS classification (2010) 02.30.Zz; 03.65.-w; 03.65.Ge; 03.65.Nk. Mathematics Subject Classification (2010): 34L25; 34L40; 81U05; 81Uxx. }}
\author{ Ricardo Weder\thanks {Fellow, Sistema Nacional de Investigadores.}\thanks{ Electronic mail: weder@unam.mx. Home page: http://www.iimas.unam.mx/rweder/rweder.html} \\
Departamento de F\'{\i}sica Matem\'atica.\\
 Instituto de Investigaciones en Matem\'aticas Aplicadas y en
 Sistemas. \\
 Universidad Nacional Aut\'onoma de M\'exico.\\
  Apartado Postal 20-126,
M\'exico DF 01000, M\'exico.}

\maketitle

\vspace{.5cm}
 \centerline{{\bf Abstract}}
  We prove a bound, of Bargmann-Birman-Schwinger type, on the number of eigenvalues of  the matrix Schr\"odinger operator on the half line, with the most general self adjoint boundary condition at the origin, and with  selfadjoint matrix  potentials that are integrable and have  a finite first moment.

\bigskip

\section{Introduction}\sss
In this paper we study the matrix Schr\"odinger operator on the half line
\begin{equation}
H_{A,B}\psi:= -\psi''+V(x)\,\psi,\qquad x\in(0,\infty),\label{1.1}
\end{equation}
where the prime denotes the derivative with respect to
the spatial coordinate $x$. The wavefunction $\psi(x)$ 
will be either a column vector with $n$
components or  an $n\times n$ matrix-valued function. As it is well known,  the most general selfadjoint boundary condition at $x=0$ for the operator \eqref{1.1} can be formulated in several equivalent way, see \cite{11}-\cite{wtf}.  However, it was proved in \cite{19}-\cite{wtf} that, without losing generality, it is useful  to state them  in terms of constant $n\times n$ matrices $A$ and $B$ as follows, 
\begin{equation}-B^\dagger\psi(0)+A^\dagger\psi'(0)=0,\label{1.2}
\end{equation}
\begin{equation}-B^\dagger A+A^\dagger B=0,\label{1.3}\end{equation}
\begin{equation}A^\dagger A+B^\dagger B>0.\label{1.4}\end{equation}
Remark  that  $A^\dagger B$ is selfadjoint.

In what follows we suppose that  the potential  $V$ is  a $n\times n$ selfadjoint matrix-valued function 

\begin{equation}V(x)=V(x)^\dagger,\qquad x\in{\bf R}^+.
\label{1.6}
\end{equation}
By  the dagger we designate the matrix adjoint .

Furthermore, we assume that  $V(x)$   belongs to the Faddeev class,  that is to say,  each entry of the matrix $V$ is Lebesgue measurable on $(0, \infty)$ and,

\begin{equation}\int_0^\infty
dx\,   (1+x)\, ||V(x)||<+\infty.
\label{1.7}
\end{equation} 
Here,  $||V(x)||$ designates  the norm of  $V(x)$ as an operator on $\mathbf C^n$. Of course,  \eqref{1.7}  holds
if and only if it holds for each entry of  $V$.

As usual, we denote by $L^2$ the standard Hilbert space of measurable functions  defined on $(0,\infty)$ with values in $\mathbf C^n$.  

It is proven in \cite{wst} (see also Section 3 below)  that the formal  differential operator \eqref{1.1} has a selfadjoint, bounded below,  realization  in $L^2$, defined by quadratic forms,  with the boundary condition  \eqref{1.2}. We also denote this selfadjoint realization  by $H_{A,B}$. For this purpose, we only need that \eqref{1.6} holds and that the potential matrix is integrable,

\begin{equation} \label{1.8}
\int_0^\infty
dx\,  ||V(x)||<+\infty.
\end{equation} 
 Furthermore, it is proven in  \cite{wst}  that $H_{A,B}$ has no singular continuous spectrum and its  absolutely continuos spectrum is $[0,\infty)$. Moreover, $H_{A,B}$ has no positive or zero eigenvalues and its negative eigenvalues are of finite multiplicity and can only accumulate at zero. If $V$ is in the Faddeev class \eqref{1.7} the number of eigenvalues of $H_{A,B}$ is finite \cite{21}, \cite{wst}.      
 
 Note that the  matrices $A, B$  in (\ref{1.2}. \ref{1.4}) are not uniquely defined. We can multiply them on the right by an invertible matrix $T$ without affecting  \eqref{1.2}, \eqref{1.3} and \eqref{1.4}, and furthermore,
\beq\label{1.5}
H_{A\,T, B\,T}= H_{A,B}.
\ene
Let $U$ we the following matrix
\beq \label{1.8b}
U:= \left( B-iA \right)\, \left( B+ i A \right)^{-1}.
\ene
In proposition 4.2 of \cite{21} it is proven that $B+iA$ is invertible and that $U$ is unitary.  Clearly, $U$ is invariant under the transformation $ (A,B) \rightarrow (A T, B T)$ for any invertible matrix $T$.  Let $ M$ be a unitary matrix that diagonalizes $U$,

\beq \label{1.9}
 M^{\dagger} \, U \,  M = \textrm{diag}\left\{ e^{2i \cos\theta_1}, e^{2i\cos\theta_2}, \cdots e^{2i\cos\theta_n} \right\},
\ene
where $ 0 < \cos\theta_j \leq \pi$.   
In the general case   there 
are $n_{\mathrm N}$ values with $\theta_j=\pi/2$ and $n_{\mathrm D}$ values with $\theta_j=\pi,$
and in consequence  there are $n_{\mathrm M}$ remaining values,
with $n_{\mathrm M}:=n-n_{\mathrm N}-n_{\mathrm D},$ such that the corresponding
$\theta_j$-values lie in
the interval $(0,\pi/2)$ or $(\pi/2,\pi).$We allow for  the special cases where any of $n_{\mathrm N},$ $n_{\mathrm D},$ and $n_{\mathrm M}$ may be zero or $n.$ The subscripts N, D, and M  refer, respectively, to Neumann, Dirichlet or mixed boundary  conditions. In fact, in the representation where the matrices $A,B$ are diagonal $\theta_j= \pi/2$ corresponds to  the Neumann boundary condition, $\theta_j= \pi$  to the Dirichlet boundary condition and $\theta$ in  $(0,\pi/2)$ or $(\pi/2,\pi)$ to mixed boundary conditions. 

See Section 2 for these issues.

Let us decompose $V$ into its positive and negative parts, i.e.,

\beq \label{1.10}
V(x)=  V_+(x) - V_-(x),\, \textrm{with} \, V_{\pm}(x) \geq 0.
\ene

By $ V_\pm (x) \geq 0$ we mean that the matrices $V_\pm(x)$ have nonnegative eigenvalues, or equivalently,  that they are nonnegative operators on $ \mathbf C^n$.

We denote by $n_{M,\textrm{b}}$ the number of $ \theta_j$ with $  0 <  \theta_j  <\pi/2.$ We designate by $ \Theta_T$ the following diagonal  $ n \times n$ matrix,
\beq \label{1.11}
\Theta_T:= \textrm{diag}\left\{ \widehat{\tan\theta_1}, \widehat{\tan\theta_2}, \cdots, \widehat{\tan\theta_n}  \right\},
\ene
where $ \widehat{\tan \theta_j}= 0$ if $ 0 <\theta_j \leq \pi/2$, $\widehat{\tan\theta_j}= \tan\theta_j$ if $ \pi/2 <  \theta_j \leq  \pi,$   for $j=1,2,\cdots, n$.

Our Bargmann-Birman-Schwinger bound \cite{bar}, \cite{bi}, \cite{sch} is the following theorem.

\begin{theorem} \label{theor}
Suposse that the boundary conditions are given by  \eqref{1.2} where the matrices $A,B$ satisfy \eqref{1.3}, \eqref{1.4} and the matrix potential satisfies (\ref{1.6}, \ref{1.7}. Let us denote by $N_{A,B}$ the number of negative eigenvalues of $H_{A,B}$. Then,
\beq \label{1.12}
N_{A,B}\leq n_{M,\textrm{b}}+  n_N + \int_0^\infty\, \hbox{\rm trace}\,  \left[V_-(x)\,(x- M\, \Theta_T\, M^\dagger )\right]\, dx.
\ene
\end{theorem}
Note that it is necessary to have $n_{M,\textrm{b}}$ and $n_N$ in the right-hand side of \eqref{1.12}, see Remark \ref{remark}.

There is currently a considerable interest on this problem.  Matrix Schr\"odinger operators on the half line are important in quantum graphs, in quantum wires
and  in quantum mechanical scattering of particles with internal structure. See, for example, \cite{11}- \cite{10} and \cite{5}- \cite{18},  and the references quoted there.

The paper is organized as follows.
In Section~2 we introduce notations and definitions and we state preliminary results that we need. In Section~3, following \cite{wst}, we define the matrix Schr\"odinger operator as a selfadjoint operator by means of quadratic form techniques and we  estimate the difference in the number of bound states of two  matrix Schr\"odinger operators with the same potential  and with different boundary conditions. In Section 4 we  state results from \cite{wst} on the integral kernel of the resolvent in the case where the potential matrix is identically zero. In Section 5 we prove our Birman-Schwinger bound on the number of bound states.

\section{Notations, definitions and preliminary results}\sss

We designate  by  ${\bf C}^+$ the upper-half complex plane,  by ${\bf R}$  the real axis, and we let ${\overline{{\bf C}^+}}:={\bf C}^+\cup{\bf R}$.  For any $k \in  \overline{{\bf C}^+}$ we denote by $k^*$ its complex conjugate. For any matrix $D$ we designate by $D^\dagger$ its adjoint. 


By $H_l, l=1,2, $ we denote  the Sobolev space of order $l$ of all square integrable, complex valued functions defined in $(0, \infty)$  with all distributional derivatives up to order $l$ given by square integrable functions \cite{24}. We designate by $H_{1,0}$ the completion of $C^\infty_0((0, \infty))$ in the norm of  $H_1$, where $C^\infty_0((0, \infty))$ is the space of all infinitely differentiable complex valued functions with compact support. We use the notation,
\beq \label{2.1}
\mathbf{H}_l:=\oplus_{j=1}^n\, H_{l}, \quad l=1,2,
\ene
for the first and second Sobolev spaces of functions with values in $\mathbf C^n$.

For any trace class operator $G$  we denote by  $\textrm{trace}(G)$ its trace. For any densely defined operator $D$ in a Banach space we denote by $\rho(D)$ its resolvent set, i.e., the open set of all $ z \in \mathbf C$ such that $D-z$ is invertible  and $\left( D-z\right)^{-1}$ is bounded. We denote the resolvent of $D$ by $R_D(z):= \left(D-z\right)^{-1}$ for $z \in \rho(D)$ \cite{ka}.


Motivated by the general selfadjoint boundary condition \cite{26,25,27} in the scalar case, i.e. when  $n=1$, we  consider the case where the matrices $A,B$ are diagonal.  We denote this special pair of diagonal matrices by  $\tilde A$ and $\tilde B,$ where

\begin{equation}\tilde {A}:=-\mathrm{diag}\{\sin\theta_1,\dots,\sin\theta_n\},
\quad \tilde B:=\mathrm{diag}\{\cos\theta_1,\dots,\cos\theta_n\}.\label{3.25}
\end{equation}
For these matrices  the boundary conditions \eqref{1.2} are,
\beq\label{3.25b}
\cos\theta_j\, \psi_j(0)+ \sin\theta_j\, \psi_j'(0)=0, \qquad j=1,2,\cdots,n.
\ene

The real parameters $\theta_j$ take values in the interval $(0,\pi].$
The the case $\theta_j=\pi/2$ corresponds to the Neumann boundary condition, case $\theta_j=\pi$ corresponds to the Dirichlet boundary condition,  and the case where $ \theta_j \neq  \pi/2, \pi$ corresponds to mixed boundary conditions. We suppose that there
are $n_{\mathrm N}$ values with $\theta_j=\pi/2$ and $n_{\mathrm D}$ values with $\theta_j=\pi,$
and hence there are $n_{\mathrm M}$ remaining values,
with $n_{\mathrm M}:=n-n_{\mathrm N}-n_{\mathrm D},$ such that the corresponding
$\theta_j$-values lie in
the interval $(0,\pi/2)$ or $(\pi/2,\pi).$
The special cases where any of $n_{\mathrm N},$ $n_{\mathrm D},$ and $n_{\mathrm M}$ may be zero or $n$ are allowed. 
Observe  that $\tilde{A}, \tilde{B}$ satisfy  \eqref{1.3}, \eqref{1.4}.

In Proposition 4.3 of \cite{21} it is proven that for any pair of matrices $(A,B)$ that satisfy \eqref{1.2}-\eqref{1.4} there is a pair of diagonal matrices $(\tilde{A}, \tilde{B})$ as in \eqref{3.25}, a unitary matrix $M$ and a two invertible matrices $T_1,T_2$ such that,
\beq \label{3.26}
A=    M\, \tilde{A}\, T_2 \, M^\dagger\, T_1, \quad  B=    M\, \tilde{B}\, T_2 \, M^\dagger\, T_1.
\ene
Note that $ T_1,T_2$ in \eqref{3.26} correspond, respectively to  $  T_1^{-1},T_2^{-1}$ in Proposition 4.3 of \cite{21}. Furthermore (see the proof of Proposition 4.3 of \cite{21})
the $ \theta_j, j=1,2,\cdots,n$  in \eqref{3.25} coincide with $\theta_j, j=1,2,\cdots,n$ that appear in the diagonal representation of the matrix $U$ in \eqref{1.9}.  

 We  consider some $n\times n$ matrix solutions to the equation  
 \begin{equation}
 -\psi''+V(x)\,\psi = k^2\, \psi,\qquad x\in(0,\infty), k \in \overline{{\bf C}^+},
 \label{3.1}
\end{equation}
assuming that $V$ satisfies \eqref{1.6}, \eqref{1.8}.

The Jost solution ( see \cite{20})  to \eqref{3.1}  is the $n\times n$ matrix solution
satisfying, for $k\in{\overline{{\bf C}^+}}\setminus\{0\},$ the asymptotics

\begin{equation}f(k,x)=e^{ikx}[I_n+o(1/x)],\quad
f'(k,x)=ik\,e^{ikx}[I_n+o(1/x)],\qquad x\to+\infty,\label{3.2}
\end{equation}
where $I_n$ denotes the $n\times n$ identity matrix. It is well  is known \cite{19,20}, that for each fixed $x,$ 
$f(k,x)$ and $f'(k,x)$ are analytic for $k\in{\bf C}^+$
and continuous for $k\in{\overline{{\bf C}^+}}$.
It follows from \eqref{3.2}   that for each fixed $k \in{\bf C}^+$, each of the $n$ columns of
$f(k,x)$  decays exponentially to zero as $x\to+\infty$.

The matrix Schr\"odinger equation \eqref{3.1} also has the $n\times n$ matrix solution $g(k,x)$
that satisfies , for each $k\in{\overline{{\bf C}^+}}\setminus\{0\},$ the following asymptotics  \cite{20}

\begin{equation} \label{3.5}
g(k,x)=e^{-ikx}[I_n+o(1/x)],\, g'(k,x)= -i k \, e^{-i k x}\, (I+ o(1/x)),  \quad x \rightarrow \infty.
\end{equation} 

It is proven in  \cite{20} that $g(k,x)$ and $g'(k,x)$ are analytic in $k\in{\bf C}^+$
and continuous in $k\in{\overline{{\bf C}^+}}\setminus\{0\}$ for each fixed $x.$ Equation \eqref{3.5} implies  that each of the $n$ columns of $g(k,x)$  grows exponentially as $x\to+\infty$ for each fixed $k\in{\bf C}^+.$

 On page  28 of \cite{20} it is  proven that  for each
$k\in{\overline{{\bf C}^+}}\setminus\{0\},$ the combined $2n$ columns of $f(k,x)$ and of $g(k,x)$
form a fundamental set of solutions to  \eqref{3.1}. Hence,
any column-vector solution $\omega(k,x)$ to  \eqref{3.1}  can be written as a linear combination of them,

\begin{equation}\omega(k,x)=f(k,x)\,\xi+g(k,x)\,\eta,\label{3.7}
\end{equation}
for some constant column vectors $\xi$ and $\eta$ in ${\bf C}^n$.

Another important $ n \times n$ matrix solution to \eqref{3.1} is the  regular solutions, $\varphi_{A,B}(k,x)$ that satisfies the initial conditions,
\begin{equation}\label{3.7b}
\varphi_{A,B}(k,0)=B, \quad \varphi'_{A,B}(k,0)=A.
\ene

\section{The matrix Schr\"odinger operator} \sss
Here we follow \cite{wst} where details and proof are given.
\subsection{The case of zero potential }
See  Subsection 4.1 of \cite{wst}.
We denote by $H_{0, A,B}$ the self- adjoint  realization of
$ - \frac{d^2}{ d x^2} $ with the boundary condition \eqref{1.2}, namely
\beq\label{10.1}
H_{0,A,B} \psi= -\frac{d^2}{d x^2} \,\psi, \qquad \psi \in D(H_{0,A,B}),
\ene
where
\beq\label{10.2}
D(H_{0,A,B}):= \left\{ \psi \in \mathbf{H}_2 : -B^\dagger\,\psi(0)+A^\dagger\,\psi'(0)=0 \right\}.
\ene

Note that $H_{0,A T,B T}= H_{0,A,B}$ for all invertible matrices $T$. Recall that in the particular case of the diagonal matrices $ \tilde{A}, \tilde{B}$ \eqref{3.25} the boundary conditions \eqref{1.2} are given by \eqref{3.25b}. These equations  can be written as,
\beq \label{10.4}
\psi'(0)= - \cot\theta_j\, \psi_j(0), \,\mathrm{if} \,\theta_j \neq \pi, \,\mathrm{and} \qquad \psi_j(0)=0, \,\mathrm{if} \,\theta_j= \pi.
\ene

Let us construct the quadratic form associated to $ H_{0, \tilde{A}, \tilde{B}}$. We denote,
\beq \label{10.5}
H_{1,j}:= H_{1,0},\, \mathrm{if}\, \theta_j= \pi, \, \mathrm{and}\, \qquad H_{1,j}:= H_1,\, \textrm{if}\, \theta_j \neq \pi.
\ene
We designate,
\beq \label{10.6}
\mathbf{H}_{1,\tilde{A}, \tilde{B}}:= \oplus_{j=1}^n\, H_{1,j}.
\ene
We define the quadratic form with domain $\mathbf{H}_{1, \tilde{A},\tilde{B}}$,

\beq\label{10.7}
h_{0, \tilde A, \tilde B}(\varphi,\psi):= \left(\varphi', \psi'\right) -\sum_{j=1}^n \widehat{\cot} \theta_j\, \varphi_j(0) \,\overline{\psi_j(0)},
\ene
where $ \widehat{\cot} \theta_j=0$ if $ \theta_j=\pi/2$, or $\theta_j= \pi$, and $ \widehat{\cot} \theta_j= \cot \theta_j$ if $\theta_j \neq \pi/2, \pi$.

The symmetric form $h_{0,\tilde A,\tilde B}$ is closed and bounded below. It follows from Theorems 2.1 and 2.6 in chapter 6 of \cite{ka} that  $H_{0, \tilde{A}, \tilde{B}}$ is the selfadjoint bounded below operator associated to the quadratic form $ h_{0 \tilde A, \tilde B}$.

We define the diagonal matrix

\beq\label{10.8}
\Theta:= \mathrm{Diag}\,\{\widehat{\cot} \theta_1, \widehat{\cot} \theta_2, \cdots \widehat{\cot}\theta_n \}.
\ene
The quadratic form associated to $H_{0,A,B}$ is given by,

\beq\label{10.9}
h_{0,A,B}\left(\varphi,\psi\right):= \left( \varphi',\psi' \right)- \sum_{j=1}^n \left< M \Theta M^\dagger  \varphi(0), \psi(0)      \right>,
\ene 
where by $ <\cdot,\cdot>$ we denote the scalar product in $\mathbf C^n$, and the domain of $h_{0,A,B}$ is given by

\beq \label{10.10}
D\left( h_{0,A,B}\right)= \mathbf{H}_{1,A,B}\, \qquad \textrm{where}\,\,      \mathbf{H}_{1,A,B}:=  M \mathbf{H}_{1,\tilde{A}, \tilde{B}} \subset \mathbf{H}_1.
\ene

\subsection{The case of integrable potential}
See Subsection 4.2 of \cite{wst}.
Suppose that $ V$ satisfies \eqref{1.6}, \eqref{1.8}. Let us define the following quadratic form,

\beq \label{10.11}
h_{A,B}\left(\varphi,\psi\right):= h_{0,A,B}\left( \varphi,\psi\right)+ \left(V\varphi,\psi\right), \qquad D(h_{A,B})= \mathbf{H}_{1,A,B}.
\ene
The symmetric form $h_{A,B}$  is closed,  and bounded below. Let us denote by 
$H_{A,B}$ the associated bounded below selfadjoint operator (see theorems 2.1 and 2.6 in chapter 6 of \cite{ka} ) . Note that \cite{wst}

$$
D\left( H_{A,B}  \right)= \left\{ \psi \in  \mathbf{H}_{1,A,B} : -B^\dagger\psi(0)+A^\dagger\psi'(0)=0,    - \psi'' + V \psi  
\in L^2 \right\}.
$$
Actually, since  $ D\left( H_{A,B} \right) \subset  \mathbf{H}_{1,A,B}$, by Sobolev's theorem, any  function $\psi$ in the domain of $H_{A,B}$ is bounded. Denote,   $\phi=  - \psi'' + V \psi$. As $\psi$ is bounded and $V$  satisfies \eqref{1.8}, $ V\psi $ is integrable on $[0,\infty)$.  Furthermore, as $\phi \in L^2$, we have that $ \phi$ is integrable on $[0,R]$ for any $ R >0$. Then, $ \psi''= -\phi + V \psi$ is integrable  on $[0,R]$ for any $ R >0$. It follows that $ \psi$ and $  \psi'$ are absolutely continuous on $[0,\infty]$ and then,  $ \psi(0)$ and $\psi'(0)$ are well defined. 

It  proven in \cite{wst} that under the  transformation \eqref{3.26}

\beq \label{3.24}
H_{ V, A, B}= M \, H_{ M^\dagger V M ,\tilde{A}, \tilde{B}}\, M^\dagger,
\ene
where we made explicit the dependence in $V$ of the matrix Schr\"odinger operator. Actually, this follows from the fact that $H_{A,B}$ is the selfadjoint operator associated  to the quadratic form $h_{A,B}$ and from the uniqueness of the selfadjoint operator associated to a quadratic form (Theorem 2.1 and Theorem 2.6 of \cite{ka}).

We wish to estimate the difference in the number of bound states of two  matrix Schr\"odinger operators with the same potential  and with different boundary conditions.
For simplicity we assume that $V$ satisfies  \eqref{1.6}, \eqref{1.8}, that it is bounded and that $V(x)=0, 0 \leq 0 \leq \delta$ for some $ \delta >0$. In this case $H_{A,B}$ is just the operator sum of $H_{0,A,B}$ and $V$ and $D(H_{A,B})= D(H_{0,A,B}).$ Since the number of bound states is invariant under unitary transformations by \eqref{3.24} it is enough to study  the case where the boundary matrices are diagonal. So,  let us consider two sets of diagonal matrices by  $(\tilde {A}_1, \tilde {B}_1)$ and $(\tilde {A}_2,\tilde{ B}_2)$ where

\begin{equation}\label{m.1}
\tilde {A}_j:=-\mathrm{diag}\{\sin\theta_{j,1},\dots,\sin\theta_{j,n}\},
\quad \tilde{B}_j:=\mathrm{diag}\{\cos\theta_{j,1},\dots,\cos\theta_{j,n}\}, \quad j=1,2.
\end{equation}
Let  $L:=\{ l_1,l_2,\cdots, l_q \}$  where $1 \leq  q  \leq n$ be  a subset of $\{1,2,\cdots,n \}$,  
and assume that  $ \theta_{1,l_m }= \theta_{2,l_m}, m=1,2,\cdots, q$.
We denote,

\beq\label{m.2} \begin{array}{c}
D_L :=  \left\{ \varphi \in C^\infty ([0, \infty), \mathbf C^n): \,\textrm{For}\,  m=1,2,\cdots, q,\varphi_{l_m}\, \textrm{has compact support in }\, [0,\infty),
  \textrm{and}
 \right.  \\\\
 \left.
\cos\theta_{j,l_m} \, \varphi_{l_m}(0)+ \sin\theta_{j,l_m}\, \varphi_{l_m}'(0)=0.
 \textrm{  For } 
m \in \{1,2,\cdots, n\} \setminus \{ l_1, l_2, \cdots, l_q \}, \right. \\\\
\left.  \varphi_m \, \textrm{has compact support in }\, (0,\infty)\right\}.
\end{array}
\ene

Let us denote by   $h_0$ the symmetric operator that acts as $ -\Delta +V$ on the domain $D(h_0)=  C^\infty_0(0,\infty),$ and let  us designate  by   $h_L$ the symmetric operator that acts as $ -\Delta +V$ on the  domain $D(h_L)= D_L.$ Note that   $H_{A_j,B_j}, j=1,2$ are selfadjoint extensions of $h_L$.

 Let us compute the deficiency indices of $h_L$. We  denote by $h$ the adjoint of $h_0$. By Theorem 3.6 of \cite{wei} and since $V$ is bounded,
 
 $$
 D(h)= \left\{ \varphi \in L^2: \varphi\, \textrm{and}\, \varphi' \, \textrm{are absolutely continuous in }\, (0,\infty)\, \textrm{and}\, \varphi'' \in L^2 \right \}. 
 $$
Since $ h_0 \subset h_L$ we have that, $h_L^\ast \subset h$. As by  Theorem 3.2 of \cite{wei} if $ \psi \in D(h), \psi$ and $\psi'$ are continuous a $x=0$, it follows integrating by parts that,
$$
\left( h_L \varphi,  \psi\right)- \left( \varphi, h_L^\ast \psi  \right)= \varphi'(0)\, \overline{\psi(0)}- \varphi(0)\, \overline{\psi'(0)}, \forall \varphi \in D(h_L),  \forall\psi \in D(h_L^\ast). 
$$
It follows that  $ \cos \theta_{j,l_m} \psi_{l_m}(0)+ \sin\theta_{j,l_m} \psi_{l_m}'(0)=0, m=1,2,\cdots, q$ for all $\psi \in D(h_L^\ast).$ 

To compute the deficiency indices of $h_L$  we have to calculate the number of linearly independent solutions of,
\beq \label{m.eq}
(-\Delta +V) \varphi^\pm = \pm i \, \varphi^\pm, \textrm{with}\, \varphi^\pm \in D(h)
\ene
such that 
\beq \label{m.3}
\cos\theta_{j,l_m} \, \varphi^\pm_{l_m}(0)+ \sin\theta_{j,l_m}\, \varphi^{\pm '}_{l_m}(0)=0, m=1,2,\cdots,q.
\ene
By \eqref{3.7} and since $ \varphi^\pm \in L^2$, 
$$
\varphi^\pm= f(\sqrt{\pm i}, x) \,v^\pm, \quad \textrm{for some}\, v^\pm \in \mathbf C^n. 
$$
But, since $ V(x)=0, 0 \leq x \leq  \delta,$ for some $ \delta >0$,
$$
f(\sqrt{\pm i},x)_{l,j} =  f_l(\sqrt{\pm i}, x)\, \delta_{l,j}, l,j =1,\cdots,n, 0 \leq x  \leq \delta,
$$
where $   f_l(\sqrt{\pm i}, x)=  \left(a_{\pm, l} \,e^{i \sqrt{\pm i} x}+ b_{\pm, l}\,  e^{- i \sqrt{\pm i} x}\right)$   for some complex numbers $a_{\pm, l}, b_{\pm, l}, l=1,2,\cdots, n$. 
Then,
$$
\cos\theta_{j,l_m} \, \varphi^\pm_{l_m}(0)+ \sin\theta_{j,l_m}\,  \varphi^{\pm '}_{l_m}(0)= \left(\cos\theta_{j,l_m}  f_{l_m}(\sqrt{\pm i},0 ) + \sin\theta_{j,l_m}  f_{l_m}'(\sqrt{\pm i},0 )
\right) v^\pm_{l_m}.
$$
We prove that   $\cos\theta_{j,l_m}  f_{l_m}(\sqrt{\pm i},0 ) + \sin\theta_{j,l_m}  f_{l_m}'(\sqrt{\pm i},0 ) \neq 0$ for $ m=1,2,\cdots,q$.  Suppose, on the contrary, that some
 $\cos\theta_{j,l_m}  f_{l_m}(\sqrt{\pm i},0 ) + \sin\theta_{j,l_m}  f_{l_m}'(\sqrt{\pm i},0 ) = 0$. Take  $\xi^\pm= (\xi^\pm_1,\xi^\pm_2, \cdots,\xi^\pm_n) \in \mathbf C^n$ with $\xi^\pm_{l_m}=1$ and $\xi^\pm_l=0, l \neq l_m$. Then, $ f(\sqrt{\pm i},x)\xi^\pm \in L^2$ will satisfy the boundary condition  $ -\tilde{B}^\dagger\,  f(\sqrt{\pm i},0) \xi^\pm + \tilde{A}^\dagger\,  f'(\sqrt{\pm i},0) \xi^\pm=0$ and this is not possible since as $H_{\tilde{A}, \tilde{B}}$ is selfadjoint it can not have complex eigenvalues.  
 
Then, \eqref{m.3} implies that $ v^\pm_{l_m}=0, m=1,2,\cdots q.$ Hence, as it follows from \eqref{3.2} that the columns of $ f(k,x)$ are linearly independent solutions of \eqref{3.1} the number of linearly independent solutions $ \varphi^\pm$ of   \eqref{m.eq} that satisfy \eqref{m.3} is $n-q$, and it follows that both deficiency indices of $h_L$ are equal to $n-q$. 

Since $H_{\tilde{A}_j, \tilde{B}_j}, j=1,2$ are selfadjoint extensions of $h_L$ it follows from Lemma 2 and Theorem 3 in Section 3 of Chapter 9   of \cite{bz} that,
\beq \label{m.4}
N_{\tilde{A}_1,\tilde{B}_1} - q \leq  N_{\tilde{A}_2, \tilde{B}_2} \leq    N_{\tilde{A}_1, \tilde{B}_1} + q.
\ene
\section{The  resolvent with zero potential}\sss
See Subsection 5.1 of \cite{wst} for the results below.
We first consider the case of the diagonal matrices $\tilde{A}, \tilde{B}$ given in \eqref{3.25}. Let us denote by $ R_{0, \tilde{A}, \tilde{B}}(z)$ the resolvent of $H_{0,\tilde{A}, \tilde{B}}$,

 Let, $R_{0, \tilde{A}, \tilde{B}, }(z)(x,y)$ be the integral kernel of  $R_{0,\tilde{A}, \tilde{B}}(z)$. Then, we have that,

\beq \label{11.4}
R_{0, \tilde{A}, \tilde{B}}(z)(x,y)=\begin{cases} \varphi_{0,\tilde{A}, \tilde{B}}(x,k)\, e^{ik y} \,\left[J_{0, \tilde{A},\tilde{B}}(k)\right]^{-1} , \quad x \leq y, \\
e^{ik x}\,   \, \varphi_{0,\tilde{A}, \tilde{B}}(y,k)\,   \left[J_{0, \tilde{A}, \tilde{B}}(k)\right]^{-1}    , \quad x \geq y, 
\end{cases}
\ene
where, $k:= \sqrt{z}, \textrm{Im}\, k \geq 0$,  $\varphi_{0,\tilde{A},\tilde{B}}(k,x)$ is the regular solution \eqref{3.7b} with zero potential,
\beq\label{11.4b}
\varphi_{0, \tilde{A}, \tilde{B},j}= \begin{cases} \displaystyle-\frac{1}{k}\sin kx,  \,\textrm{if}\, \theta_j=\pi, \\
\displaystyle-\cos kx, \, \textrm{if}\, \theta_j= \pi/2, \\
\displaystyle\frac{1}{k} \cos\theta_j\, \sin kx -\sin\theta_j\, \cos kx,  \,\textrm{if} \,\theta_j \neq \pi, \pi/2,
\end{cases}
\ene
and   and $J_{0,\tilde{A}, \tilde{B}}$ is the Jost matrix,
\begin{equation} J_{0, \tilde{A}, \tilde{B}}(k)=
\mathrm{diag}\left\{ \cos \theta_1+i k \sin\theta_1,\dots,
 \cos\theta_{n_{\mathrm M}}+i k \sin\theta_{n_{\mathrm M}},-I_{n_{\mathrm D}}, ik\,I_{n_{\mathrm N}}\right\},
\label{3.29}
\end{equation}

We designate by $ R_{0, A, B}(z)$ the resolvent of $H_{0,A,B}$. Then, by  \eqref{3.26}, \eqref{3.24}, 
\beq\label{11.5}
R_{0, A, B}(z)= M \,  R_{0, \tilde{A}, \tilde{B}}(z)\, M^\dagger, \quad z \in \rho \left( H_{0,A,B} \right).
\ene
Its integral kernel is given by,
\beq \label{11.6}
R_{0, A, B}(z)(x,y)=\begin{cases}\varphi_{0,A, B}(x,k)\, e^{ik y}\, \left[J_{0, A,B}(k)\right]^{-1}, \quad x \leq y, \\
 e^{ik x}\,   \varphi_{0,A, B}(y,k) \,  \left[J_{0, A, B}(k)\right]^{-1}, \quad x \geq y. 
\end{cases}
\ene
 where  $J_{0,A,B}= B-ikA$ is the Jost matrix. The following estimate holds \cite{wst}, 
\beq\label{11.6b}
\left|  R_{0, A, B}(z)(x,y) \right| \leq C  D(k) \, e^{-\textrm{Im} k |x-y|}, D(k):= \max \left[ \frac{1}{|k|},
\frac{1}{\left|\cos \theta_1+ik\sin\theta_1\right|},\dots,
\frac{1}{\left|\cos\theta_{n_{\mathrm M}}+ik\sin\theta_{n_{\mathrm M}}\right|}\right],
\ene
where $ k=\sqrt{z}$

\section{The number of eigenvalues}
\sss
Let us first consider the case of zero potential. It is convenient  to go to the diagonal representation where the boundary matrices  are given by \eqref{3.25} and the boundary conditions by  \eqref{3.25b} and with $n_N, n_D,$ and $n_M$ defined below \eqref{3.25b}. It follows from a simple calculation that the $n_N$ Neumann boundary conditions and the $n_D$ Dirichlet boundary conditions give rise to no eigenvalues. Also the mixed boundary conditions with   $\tan \theta_j < 0$ produce no eigenvalues. However,  the mixed boundary conditions with $\tan \theta_j >0$ produce the eigenvalue $-(\cot\theta_j)^2$. To deal with this issue we will take advantage of the fact that changing  $q$ boundary conditions can only add $q$ eigenvalues, to turn the mixed boundary conditions with   $\tan \theta_j >0$     into boundary conditions with  $\tan \theta_j <0$. We will also find convenient to turn the Neumann boundary conditions into mixed boundary conditions with  $\tan \theta_j <0$,  to avoid a singularity at zero energy that appears in the  integral kernel of the resolvent with zero potential \eqref{11.4}   when there are Neumann boundary conditions present              
For any bounded below selfadjoint operator $H$ let us denote
\beq \label{mm}
\mu_n(H):= \sup_{\varphi_1,\varphi_2,\cdots,\varphi_{n-1}}\left( \inf_{\{\varphi \in Q(H), \|\varphi\|=1, \varphi \,\textrm{orthogonal to }\, \varphi_1, \varphi_2,\cdots,\varphi_{n-1}\}}
(H \varphi, \varphi)\right),
\ene
where $Q(H)$ is the quadratic form domain of $H$.
Then, by the min-max principle (Theorem XIII.1 of \cite{rs}), either there are $n$ eigenvalues, repeated according to multiplicity, below the  bottom  of the essential spectrum and $\mu_n(H)$ is the nth eigenvalue repeated according to multiplicity or $\mu_n$ is the bottom of the essential spectrum and $\mu_n=\mu_{n+1}= \mu_{n+2}= \cdots.$ 

Recall that under \eqref{1.6}, \eqref{1.7} the operator $H_{A,B}$ has a finite number of negative eigenvalues, it has no zero or positive eigenvalues and the essential spectrum is given by $[0,\infty)$ and it is absolutely continuous. For these results see \cite{wst}.

For any $E<  0$ we denote by $N_{A,B}(E)$ the number of eigenvalues, repeated according to multiplicity, of $H_{A,B}$ that are smaller than $E$.

Let $V_{\pm}$  be defined as in \eqref{1.10}.
Let us denote by $H_{A,B,V_-}$ the selfadjoint operator  associated to the quadratic form  \eqref{10.11}, but with $V_-$ instead of $V$.  By $N_{A,B,V_-}(E)$ we denote the number of eigenvalues, repeated according to multiplicity, of $H_{A,B,V_-}$ that are smaller than $E$. By the minimax principle, 
$$
N_{A,B}(E) \leq N_{A,B,V_-}.
$$
It follows that it is enough to prove  Theorem  \ref{theor} in the case $ V\leq 0$. 

  We have proven in \cite{wst}
that,
\beq\label{11.21} 
R_{A,B}(z)= R_{0,A,B}(z)- R_{0,A,B}(z)\, V_2  \,\left(I+ V_1 R_{0,A,B}(z)\, V_2\right)^{-1} \, V_1 R_{0,A,B}(z),
\quad z \in \rho(H_{0,A,B})\cap \rho(H_{A,b}).
\ene
where, since $ V\leq 0$,    $V_1:= \sqrt{|V|}, V_2:= - \sqrt{|V|}.$

We say that $V\in C^\infty_0(0,\infty)$ if each of the entries of $V$ is a function in 
$C^\infty_0(0,\infty)$
 Let $Q_j(x) \in C^\infty_0((0,\infty)$ be a sequence of matrix potentials such that, $ Q_j(x) \geq 0$ and 

$$
\lim_{j \rightarrow \infty} \int_0^\infty\, \sqrt{1+x}\,\left\| \sqrt{|V|}(x)-Q_j(x) \right\|^2=0.
$$
Denote 
 $$
 V_j= V_{1,j}\, V_{2,j}. \textrm{where} \, V_{1,j}:= Q_j, 
, V_{2,j}:=- Q_j.
 $$
 Let us denote by $H_{A,B,j}$ the selfadjoint operator associated to the quadratic form \eqref{10.11} with $V_j $ instead of $V$.  Note that since $V_j \in C^\infty_0(0,\infty)$  the operator $H_{A,B,j}$ is just the operator sum $H_{0,A,B}+V_j$.

By \eqref{11.6b},  the operators $V_1 \,R_{0,A,B}(z)\, V_2 $ and  $V_{1,j} \,R_{0,A,B}(z)\, V_{2,j} $ are Hilbert-Schmidt for $z \in \mathbf C^\pm$ and

$$
\lim_{j \rightarrow \infty} V_{1,j} \,R_{0,A,B}(z)\, V_{2,j}= V_{1} \,R_{0,A,B}(z)\, V_2, 
$$
where the limit is on the Hilbert-Schmidt norm. We designate by $R_{A,B,j}:= \left( H_{A,B,j}-z \right)^{-1}$. Then, by \eqref{11.21},
$$
\lim_{j \rightarrow \infty} R_{A,B,j}(z)= R_{A,B}(z), \quad z \in \mathbf C^\pm,
$$
 in operator norm. Hence, $H_{A,B,j}$ converges to $H_{A,B}$ in norm resolvent sense, and by Theorem VIII.23 in \cite{rs2}, for any $E < 0$ that is not an eigenvalue
of $H_{A,B}$
$$
\lim_{j \rightarrow \infty} P_{(-\infty, E)}(H_{A,B,j})= P_{(-\infty, E)}(H_{A,B}),
$$  
 in operator norm, where  $ P_{(-\infty, E)}(H_{A,B,j})$, respectively,  $P_{(-\infty, E)}(H_{A,B})$ denote the spectral projector for $(-\infty, E)$ of $H_{A,B,j}$ and $H_{A,B}$.
 Note that
 $$
 N_{A,B}(E)= \textrm{dim}\, P_{(-\infty, E)}(H_{A,B})\,L^2,\quad   N_{A,B,j}(E)= \textrm{dim}\, P_{(-\infty, E)}(H_{A,B,j})\,L^2.
 $$
     
 Then, by Theorem 6.32 in chapter one of \cite{ka},
  \begin{equation} \label{b.2}
 \lim_{j\rightarrow \infty}  N_{A,B,j}(E)=  N_{A,B}(E),
 \end{equation}  
 and actually the equality in   \eqref{b.2}  is obtained for a large enough finite $j$. In conclusion, it is enough to prove  \eqref{1.12} for $ V \in C^\infty_0(0,\infty)$, with $ V \leq 0$ and this what we proceed to do now. 

It is convenient  to go to the diagonal representation where the boundary matrices  are given by \eqref{3.25} and the boundary conditions by  \eqref{3.25b} and with $n_N, n_D,$ and $n_M$ defined below \eqref{3.25b}

Let us consider the operator,
\beq \label{b.3}
H_{\tilde{A},\tilde{B}}(\lambda):= H_{0,\tilde{A},\tilde{B}}+ \lambda V,
\ene
for $ 0 \leq \lambda \leq 1$. By perturbation theory \cite{ka}   the eigenvalues of $H_{\tilde{A}, \tilde{B}}( \lambda)$ are continuous functions of $\lambda$ (they are branches of holomorphic functions) and are non increasing functions of $\lambda$ because $ V \leq 0$. Then, by the mini max principle  the $\mu_n(H_{\tilde{A},\tilde{B}}(\lambda)$
are continuous non increasing functions of $\lambda$. Suppose, moreover, that there are no Neumann boundary conditions and no mixed boundary conditions with 
$ 0 < \theta_j< \pi /2$. Then, since $H_{\tilde{A}, \tilde{B},0}$ has no bound states,   $\mu_n(H_{\tilde{A},\tilde{B}}(0)=0$.

The crunch of the Birman-Schwinger \cite{bi}, \cite{sch} method is the following. By the min-max principle,
\beq \label{b.4}
N(H_{\tilde{A},\tilde{B}})(E)= \#\left\{ n:  \mu_n(H_{\tilde{A},\tilde{B}}(1) < E \right\}.
\ene 
 However, as $\mu_n(H_{\tilde{A},\tilde{B}}(0)=0$, since $H_{0,\tilde{A},\tilde{B}}$ has no eigenvalues, and $\mu_n(H_{\tilde{A},\tilde{B}}(\lambda)$ are non increasing functions,
\beq \label{b.5}\begin{array}{l}
 N(H_{\tilde{A},\tilde{B}})(E)= \#\left\{ n:  \mu_n(H_{\tilde{A},\tilde{B}}(\lambda) = 
 E, \, \textrm{for some}\, 0 < \lambda <1 \right\} \leq
 \sum_{\{ \lambda:  0 <\lambda <1 : \mu_n(\lambda)= E, n=1,2,\cdots \}}  \, \displaystyle \frac{1}{\lambda}.\\
  \end{array} \ene

Suppose that $E <0$ is an eigenvalue of $H_{\tilde{A},\tilde{B}}(\lambda )$  with eigenvector $\varphi$, i.e.,
\beq \label{b.6}
\left( H_{0,\tilde{A},\tilde{B}}+ \lambda V-E\right)\,\varphi=0. 
\ene 
Equation \eqref{b.6} holds if and only if,
\beq \label{b.7}
\frac{1}{\lambda}\,  V_1\, \varphi = V_1 \left( R_{0,\tilde{A}, \tilde{B}}(E) V_1\right) \, V_1 \varphi.
\ene
But \eqref{b.7} is equivalent to
\beq \label{b.8}
\mathcal B \psi = \frac{1}{\lambda} \psi \qquad \textrm{for some non zero }\,  \psi \in L^2,
\ene
 where $ \mathcal B$ is the operator
\beq \label{b.9} 
\mathcal B := V_1 R_{0,\tilde{A}, \tilde{B}}(E) V_1.
\ene
Note that $\mathcal B$ is a selfadjoint and non negative operator. Then $ E$ is an eigenvalue of $H_{\tilde{A}, \tilde{B}}(\lambda)$,    if and only if $1/\lambda$ is an eigenvalue of $\mathcal B$ and the multiplicity is the same. Equivalently $ \mu_n(\tilde{A}, \tilde{B}, \lambda)=E$, for some $n=1,2,\cdots$ if and only $ 1/ \lambda$ is an eigenvalue $ \mathcal B$ and the number of $n's$ such that  $ \mu_n(\tilde{A}, \tilde{B}, \lambda)=E$ is equal to the multiplicity of the eigenvalue $1/\lambda$ of $\mathcal B$.  
Denote by $ \rho_1, \rho_2, \cdots$ the eigenvalues of $\mathcal B$ repeated according to multiplicity. Hence, by \eqref{b.5}
 \beq\label{b.10}
 N(H_{\tilde{A},\tilde{B}})(E)\leq  \sum_{\{\rho_n: \rho_n>1\}} \rho_n \leq \, \textrm{trace} \,\,\mathcal B.
 \ene 
By  \eqref{11.4}     $\mathcal B$ is an integral operator with continuous kernel, $\mathcal B(x,y)$,
$$
\mathcal B(x,y):= V_1(x)\, R_{0,\tilde{A}\,\tilde{B}}(x,y)\, V_1(y).
$$
 Then (see \cite{ya1})
\beq \label{b.11}
 N(H_{\tilde{A},\tilde{B}})(E)  \leq  - \int_0^\infty  \text{trace}\left[V(x)  R_{0,\tilde{A}, \tilde{B}}(E)(x,x)\right]\, dx.
 \ene
 Taking the limit as $ E \rightarrow 0$ (note that since there are only a finite number of eigenvalues and zero is not an eigenvalue  $N(H_{\tilde{A},\tilde{B}})(E)$ is constant for $-E$ small enough ) and using \eqref{11.4}, \eqref{11.4b} and \eqref{3.29}
 \beq \label{b.12}
 N_{\tilde{A}, \tilde{B}}:=  N(H_{\tilde{A},\tilde{B}})(0) \leq  - \int_0^\infty \,\sum_{j=1}^nV_{j,j}(x)\, \left( x- \tan\theta_j \right)\, dx.
 \ene
 Let us now go back to the general case where we allow for Neumann boundary conditions and for mixed conditions with $ 0 <   \theta_j < \pi/2$. We define
  $$
  \theta_{j, +}:= \theta_j,\, \textrm{if}\, \frac{\pi}{2} <   \theta_j  \leq \pi  , \theta_{j, +}= \pi, \, \textrm{if}\,\, 0 < \theta_j \leq \frac{\pi}{2}, j=1,2,\cdots, n,
  $$
  and let  us define $\tilde {A}_+$, $\tilde{B}_+ $ as in \eqref{3.25} with $\theta_{j, +}$ instead of $\theta_j$ and,   
   $$
   H_{\tilde{A}, \tilde{B}, +}:= H_{0,\tilde{A}_+, \tilde{B}_+} + V.
   $$
   Then, by \eqref{m.4} and \eqref{b.12}
   \beq\label{b.13}
   N_{\tilde{A},\tilde{B}} \leq    n_{M, b} + n_N -  \int_0^\infty \,\sum_{j=1}^n\, V_{j,j}(x)\, \left( x- \tan\theta_{j, +} \right)\, dx,
  \ene
 where we denote by $n_{M,\textrm{b}}$ the number of $ \theta_j$ with $  0 <  \theta_j  <\pi/2$. Recall that $n_N$ is the number of Neumann boundary conditions on the diagonal representation where the boundary matrices are given by \eqref{3.25}. 
 
 
 
 Let us consider  the general case of  matrices $A,B$ that satisfyEquation \eqref{1.3}, \eqref{1.4}. Then,  equation \eqref{b.13} holds with the with the matrices $\tilde{A}, \tilde{B}$  in \eqref{3.26} and with the potential $\tilde{V}:= M^\dagger\, V M$. Finally, by \eqref{3.24}, since the number of eigenvalues is invariant under a unitary transformation, and by the cyclicity of the trace, we have that equation \eqref{1.12} holds.

\begin{remark}\label{remark}{\rm  The numbers $n_{M, \textrm{b}}$ and $n_D$ are necessary in \eqref{1.12}. Suppose that $ \tilde{V}$ is a diagonal matrix. 
$$
\tilde{V}(x)= \textrm{diag}\left\{ \tilde{V}_1(x), \tilde{V}_2(x), \cdots, \tilde{V}_n(x) \right\}.
$$ 
Then, as mentioned above, for  each $ 0 <  \theta_j < \pi /2$  produces the  eigenvalue $ -(\cot \theta_j)^2$ if the corresponding $\tilde{V}_j$ is identically zero. Then, there are $ n_{M, \textrm{b}}$  eigenvalues even if $ \tilde{V}$ is identically zero. Moreover, each $ \theta_j= \pi /2$ produces an eigenvalue if $ \tilde{V}_j= \lambda \, Q(x),$ if $ Q(x) \leq 0$  it is not identically zero, and for any $\lambda >0$. By the proof of Theorem \ref{theor}  it is enough to prove that for any $ \lambda >0$, there is an  $ E <0$  such that the following operator in $L^2(0, \infty)$, 

\beq 
\mathcal B_N:=\sqrt{-Q(x)}\, (-\Delta_N+E)^{-1}\, \sqrt{-Q(x)},
\ene
has an eigenvalue larger than $ 1/ \lambda$, where $-\Delta_N$ is the selfadjoint    realization of $ - \Delta$ in $L^2(0,\infty)$ with Neumann boundary condition at $x=0$.    
But, as $ \mathcal B_N$ is Hilbert-Schmidt it is enough to prove that,
$$
\lim_{ E \rightarrow 0}\, \left\| \mathcal B_N \right\|= \infty,
$$
and this will follow if we find a $\varphi$ so that,
\beq \label{b.15} 
\lim_{E \rightarrow 0} \left\|  \sqrt{\mathcal B_N} \varphi \right\|= \infty.
\ene 
The operator $-\Delta_N$ is diagonalized by the  cosine transform, 

$$
\hat{\varphi}(k):=  \sqrt{\frac{2}{\pi}}\, \int_0^\infty\, \cos(kx)\, \varphi(x)\, dx.
$$
Then,
$$
\lim_{E \rightarrow 0} \left\|  \sqrt{\mathcal B_N} \varphi \right\|^2= \lim_{E \rightarrow 0}\, \int_0^\infty\, \frac{1}{(k^2- E)^2}\, |\hat{\psi}(k)|^2\, dk = \infty,\text{with} \,\psi(x):= \sqrt{-Q(x)}\varphi(x),
$$ 
if we take $ \psi \in C^\infty_0(0,\infty), \psi \geq 0$, so that $ \hat{\psi}(0) >0$.   
Then, there is always a diagonal potential $\tilde{V}(x) = \lambda \, Q(x)$  such that there are $ n_N$ negative eigenvalues for any $ \lambda >0$, what shows that the integer $n_N$  is necessary in \eqref{1.12}.
}\end{remark}

\noindent{\bf Acknowledgements}

Research partially supported by project  PAPIIT-DGAPA UNAM  IN102215 and by project SEP-CONACYT CB 2015,  254062


\begin{thebibliography}{99} 
 \bibitem{11} V. Kostrykin and R. Schrader,
{\it Kirchhoff's rule for quantum wires,} J. Phys. A {\bf 32}, 595--630
(1999).

\bibitem{12} V. Kostrykin and R. Schrader,
{\it Kirchhoff's rule for quantum wires. II: The inverse problem with possible applications to quantum computers,} Fortschr. Phys. {\bf 48}, 703--716
(2000).

 \bibitem{8} M. S. Harmer, {\it Inverse scattering for the matrix Schr\"odinger
operator and Schr\"odinger operator on
graphs with general self-adjoint boundary conditions,}
ANZIAM J. {\bf 44}, 161--168 (2002).

\bibitem{9} M. S. Harmer, {\it The matrix Schr\"odinger Operator
and Schr\"odinger Operator on Graphs,} Ph.D. thesis, University of
Auckland, New Zealand, 2004.
\bibitem{10} M. S. Harmer, {\it Inverse scattering on matrices with boundary conditions,}
J. Phys. A {\bf 38}, 4875--4885 (2005).

\bibitem{19} T. Aktosun, M. Klaus, and R. Weder,
{\it Small-energy analysis for the self-adjoint matrix
Schr\"odinger equation on the half line,}
J. Math. Phys. {\bf 52}, 102101 (2011), arXiv: 1105.1794 [math-ph].

\bibitem{21} T. Aktosun, and R. Weder,
{\it High-energy analysis and Levinson's theorem for the self-adjoint matrix
Schr\"odinger operator on the half line,}
J. Math. Phys. {\bf 54}, 012108 (2013), arXiv:1206.2986 [math-ph].

 \bibitem{23} T. Aktosun, M. Klaus, and R. Weder, { \it  Small-energy analysis for the self-adjoint matrix
Schr\"odinger operator on the half line II,} J. Math. Phys. {\bf 55}, 032103 (2014), arXiv: 1310.4809 [math-ph] .

\bibitem{wst} R. Weder,  Scattering theory for the matrix Schr\"odinger operator on the half line with general boundary conditions,
 J. Math. Phys., {\bf 56}, 092103 (2015)  , arXiv:1505.01879 [math-ph].

 \bibitem{wtf} R. Weder, Trace Formulas for the matrix Schr\"odinger Operator on the half-line with general boundary conditions, J. Math.  Phys. {\bf 57},
 112101 (2016), arXiv: arXiv:1603.09432 [math-ph].

\bibitem{5} N. I. Gerasimenko,
{\it The inverse scattering problem on a noncompact graph,}
Theoret. Math. Phys. {\bf 75}, 460--470 (1988).

\bibitem{6} N. I. Gerasimenko and B. S. Pavlov,
{\it A scattering problem on noncompact graphs,}
Theoret. Math. Phys. {\bf 74}, 230--240 (1988).

\bibitem{7} B. Gutkin and U. Smilansky,
{\it Can one hear the shape of a graph?}
J. Phys. A {\bf 34}, 6061--6068 (2001).

\bibitem{17} P. Kurasov and F. Stenberg,
{\it On the inverse scattering problem on branching graphs,}
J. Phys. A {\bf 35}, 101--121 (2002).

\bibitem{13} P. Kuchment,
{\it Quantum graphs. I. Some basic structures,}
Waves Random Media {\bf 14}, S107--S128 (2004).

\bibitem{3} J. Boman and P. Kurasov,
{\it Symmetries of quantum graphs and the inverse scattering problem,}
Adv. Appl. Math. {\bf 35}, 58--70 (2005).



\bibitem{14} P. Kuchment,
{\it Quantum graphs. II. Some spectral properties of quantum and combinatorial graphs,}
J. Phys. A {\bf 38}, 4887--4900 (2005).

\bibitem{15} P. Kurasov and M. Nowaczyk,
{\it Inverse spectral problem for quantum graphs,}
J. Phys. A {\bf 38}, 4901--4915 (2005).



\bibitem{2} G. Berkolaiko, R. Carlson, S. A. Fulling, and P. Kuchment (eds.),
{\it Quantum Graphs and their Applications,} Contemporary Mathematics, 415,
Amer. Math. Soc., Providence, RI, 2006.


\bibitem{4} P. Exner, J. P. Keating, P. Kuchment, T. Sunada, and A. Teplyaev (eds.),
{\it Analysis on graphs and its applications,}
Proc. Symposia in Pure Mathematics, 77,
Amer. Math. Soc., Providence, RI, 2008.

 \bibitem{1} J. Behrndt and A. Luger,
{\it On the number of negative eigenvalues of the Laplacian on a metric graph,}
J. Phys. A {\bf 43}, 474006 (2010).


\bibitem{16} P. Kurasov and M. Nowaczyk,
{\it Geometric properties of quantum graphs
and vertex scattering matrices,}
Opuscula Mathematica {\bf 30}, 295--309 (2010).


\bibitem{18} G. Berkolaio and  P. Kuchment, {\it Introduction to Quantum Graphs}.  Mathematical Surveys and Monographs {\bf 186} Am. Math. Soc, Providence, R I 2013.

\bibitem{20} Z. S. Agranovich and V. A. Marchenko, {\it The Inverse Problem of
Scattering Theory,} Gordon and Breach, New York, 1963.












\bibitem{24} R. A. Adams, J. J. F. Fournier, {\it  Sobolev Spaces}, Elsevier Science, Oxford, U.K., 2003.

\bibitem{ka} T. Kato, {\it Perturbation Theory of Linear Operators. Second Edition}, Springer, Berlin, 1976.
\bibitem{26} V. A. Marchenko,
{\it Sturm-Liouville Operators and Applications,}
Birkh\"auser, Basel, 1986.


\bibitem{25} B. M. Levitan, {\it
Inverse Sturm-Liouville Problems,} VNU Science Press, Utrecht, 1987.

\bibitem{27} T. Aktosun and R. Weder, {\it
Inverse spectral-scattering problem with two sets of discrete
spectra for the radial Schr\"odinger equation,} Inverse Problems {\bf 22}, 89--114
(2006).



\bibitem{wei} J. Weidmann, {\it Spectral Theory of 
Ordinary Differential Operators}, Lecture Notes in Mathematics {\bf 1258} Springer, Berlin, 1987. 

\bibitem{bz} M. S. Birman, M. Z. Solomjak, Spectral Theory of Self-Adjoint Operators, D. Reidel, Dordrecht, 1987. 
 
\bibitem{rs} M. Reed and B. Simon, {\it Methods of Modern Mathematical Physics IV Analysis of Operators}, Academic Press, New York 1978. 







\bibitem{rs2}  M. Reed and B. Simon, {\it Methods of Modern Mathematical Physics I Functional Analysis }, Academic Press, New York 1972. 

\bibitem{bi} M. S. Birman, The spectrum of singular boundary problems (Russian), Mat. Sb. {\bf 55}, 125-174 (1961); Amer. Math. Soc. Transl. {\bf 53}, 23-80  (1966).  

\bibitem{sch}  J.  Schwinger, On the bound states of a given potential, Proc. Nat. Acad. Sci. U.S.A. {\bf 47}, 122-129 (1961).
 
\bibitem{ya1} D. R. Yafaev, Mathematical Scattering Theory: General Theory, Amer. Math. Soc. Providence, Rhode Island, 1992. 

\bibitem{bar} V. Bargmann, On the number of bound states on a central field of forces, Proc. Nac. Acad. Sci. {\bf 38} (11) 961-966 (1952).
 \end{thebibliography}
\end{document}